\documentclass[aps,prd,preprint,groupedaddress,superscriptaddress,nofootinbib,longbibliography,showpacs,showkeys,bibnotes]{revtex4-1}
\usepackage{graphics}%
\usepackage[mathcal]{euscript}
\usepackage[latin1]{inputenc}
\usepackage{latexsym}
\usepackage{hyperref}
\usepackage{amsmath}    
\usepackage{graphicx}   
\usepackage{verbatim}  
\usepackage{color}      
\usepackage{subfigure}    
\usepackage{bm}
\usepackage{amssymb}
\usepackage{bbm}
\usepackage{float}
\usepackage{slashed}
\usepackage{amsfonts}
\usepackage{euscript}  
\usepackage{mathrsfs} 
\usepackage{bbding}
\usepackage{pifont}
\usepackage{cancel}
\usepackage{multirow}
\usepackage{soul}
\usepackage{calrsfs}
\usepackage{dsfont}
\DeclareMathAlphabet{\pazocal}{OMS}{zplm}{m}{n}

\usepackage[usenames,dvipsnames]{xcolor}
\hypersetup{colorlinks=true, citecolor=Emerald, linkcolor=Emerald,urlcolor=Emerald}

\begin{document}

\title{Taming higher-derivative interactions and bootstrapping gravity with soft theorems}

\author{Ra\'ul Carballo-Rubio}
\email[]{raul.carballorubio@sissa.it}
\affiliation{SISSA, International School for Advanced Studies, Via Bonomea 265, 34136 Trieste, Italy}
\affiliation{INFN Sezione di Trieste, Via Valerio 2, 34127 Trieste, Italy} 
\author{Francesco Di Filippo}
\email[]{francesco.difilippo@sissa.it}
\affiliation{SISSA, International School for Advanced Studies, Via Bonomea 265, 34136 Trieste, Italy}
\affiliation{INFN Sezione di Trieste, Via Valerio 2, 34127 Trieste, Italy} 
\author{Nathan Moynihan}
\email[]{nathantmoynihan@gmail.com}
\affiliation{The Laboratory for Quantum Gravity \& Strings, Department of Mathematics \& Applied Mathematics, University Of Cape Town, Private Bag, 7701 Rondebosch, South Africa} 


\begin{abstract}
On-shell constructibility is redefining our understanding of perturbative quantum field theory. The tree-level S-matrix of constructible theories is completely determined by a set of recurrence relations and a reduced number of scattering amplitudes. In this paper, we revisit the on-shell constructibility of gravitational theories making use of new results on soft theorems and recurrence relations. We show that using a double complex shift and an all-line soft deformation allows us to relax the technical conditions for constructibility, in order to include more general propagators and higher-derivative interactions that prevent using conventional Britto-Cachazo-Feng-Witten (BCFW) shifts. From this result we extract a set of criteria that guarantee that a given gravitational action has the same tree-level S-matrix in Minkowski spacetime as general relativity, which implies the equivalence at all orders in perturbation theory between these classical field theories on asymptotically flat spacetimes. As a corollary we deduce that the scattering amplitudes of general relativity and unimodular gravity are the same for an arbitrary number of external particles (as long as the S-matrix of the latter is unitary), thus extending previous works that were able to deal only with $n=4$ and $n=5$ amplitudes.
\end{abstract}

\maketitle

\section{Introduction}\label{sec:intro}

A quantum field theory is said to be ``constructible'' \cite{Benincasa:2007xk,Schuster:2009,Cohen2010,He:2010,Cheung2015} if its amplitudes $A_n(1^{h_1}2^{h_2}...\,n^{h_n})$ for $n>n_\star\in\mathbb{Z}$ can be determined recursively by means of recurrence relations, the initial conditions (or seeds) of which are given by the set of amplitudes with $n\leq n_\star$. In constructible theories, scattering amplitudes are therefore determined by the principles that fix the form of the recurrence relations, and a reduced number of amplitudes. This represents a huge simplification with respect to the traditional calculation using Feynman diagrams, in which amplitudes have to be evaluated independently for each value of $n$, with the calculation quickly becoming cumbersome with increasing values of the latter integer (see \cite{vanEijk2004} for instance).

This higher calculation efficiency has found a large number of applications. Together with the use of spinor-helicity variables, these techniques allow arriving at particularly elegant and simple expressions for $A_n(1^{h_1}2^{h_2}...\,n^{h_n})$ (e.g., \cite{Parke:1986,Bedford:2005,Nguyen:2010}). In this paper, we exploit these methods in order to extract information regarding the physical equivalence of different theories through the complete calculation of their S-matrices. The equivalence theorem(s) states that the S-matrix is blind to (nonlinear) local field redefinitions in quantum field theory \cite{Chisholm1961,Kamefuchi1961,tHooft1973}, and so computing the full (tree-level) S-matrix of two field theories and showing their equality is a useful way to determine physical equivalence around certain classical backgrounds. Computing the full S-matrix for any number $n$ of external legs is a daunting (if not impossible) task in the traditional approach using Feynman diagrams even if restricting to tree-level processes, but constructibility makes it possible.

Moreover, we also highlight here that on-shell methods permits us to make precise statements regarding the possibility of deriving general relativity from Lorentz invariance, a classical problem that goes back to Kraichnan \cite{Kraichnan1947,Kraichnan1955}, Gupta \cite{Gupta1954,Gupta1957} and Feynman \cite{Feynman1996}. The result that general relativity is the only nonlinear theory that can be obtained from massless particles of spin 2 (gravitons, in the following) is frequently quoted in the literature but is not often scrutinized. In particular, the technical assumptions that are necessary to prove such a result (that must certainly exist) remain obscure. This is partially due to intrinsic limitations of previous analyses, that were typically off-shell and hence focused on the derivation of the Einstein-Hilbert action (in the following, we will refer to this as ``off-shell constructibility'' to distinguish it from the standard notion of on-shell constructibility). Here, we stress that modern on-shell methods provide a convenient mathematical framework for the analysis of this problem, in which these assumptions can be fleshed out. In fact, we think that this is one of the major achievements of our discussion.

The outline of this paper is as follows. We start in section \ref{sec:grcons} with a brief review of the constructibility of general relativity, mentioning some of the classic results regarding the off-shell constructibility of the Einstein-Hilbert action as well as more recent on-shell results, and highlighting (to the best of our knowledge, for the first time) their interplay. Section \ref{sec:statement} contains our main result, that determines the most general theories that have, according to their soft behavior, the same S-matrix (around flat spacetime) as general relativity. The proof of this result is given in section \ref{sec:proof}. In section \ref{sec:app} we revisit some theories of modified gravity on the light of the previous discussion. We close the paper with a brief conclusions section. 

\section{General relativity as a constructible theory \label{sec:grcons}}

Let us start with a brief summary of known results regarding the derivation of the Einstein-Hilbert action from the information encoded in the massless spin-2 representation of the Poincar\'e group \cite{Wigner1939}. First of all, it is necessary to keep in mind that one of the main goals of this approach is obtaining the features associated with diffeomorphism invariance as a consequence of a non-geometric set of principles that can be formulated entirely within the framework of quantum field theory in flat spacetime. Aside from the first works \cite{Kraichnan1947,Kraichnan1955,Gupta1954,Gupta1957,Feynman1996}, 
Deser's derivation \cite{Deser1969} seems to be the best-known approach to this problem, but we would also like to point out the thorough analysis contained in Huggins' PhD thesis \cite{Huggins1962} as well as Wald's analysis  \cite{Wald1986,Cutler1986,Wald1986b} (see also \cite{Heiderich1988,Heiderich1990}). More recent discussions include \cite{Padmanabhan2004,Butcher2009,Deser2009,Barcelo2014,Ortin2015}. The starting point is the observation that the Einstein-Hilbert action $\mathscr{S}_{\rm EH}$ can be written, up to a boundary term and performing an expansion $g_{ab}=\eta_{ab}+\kappa h_{ab}$, as
\begin{align}\label{eq:cons1}
\mathscr{S}_{\rm EH}&=\frac{1}{2\kappa^2}\int\text{d}^4x\sqrt{-g}R(g)\nonumber\\
&=\int\text{d}^4x\sqrt{-\eta}\sum_{k=2}^\infty\kappa^{k-2} \Gamma_{(k)}^{a_1b_1c_1a_2b_2c_2 i_1j_1...i_{k-2}j_{k-2}} h_{i_1j_1}\times ...\times h_{i_{k-2}j_{k-2}}\bar{\nabla}_{a_1}h_{b_1c_1}\bar{\nabla}_{a_2}h_{b_2c_2},
\end{align}
where $\kappa^2=8\pi G c^{-4}$ and $\bar{\nabla}$ is the covariant derivative associated with $\eta_{ab}$ (we are being general enough to include the possibility that coordinates other than Cartesian are used). The Einstein-Hilbert action displays specific values for the coefficients $\{\Gamma_{(k)}\}_{k=2}^\infty$. Let us note that, when going from the first line in Eq. \eqref{eq:cons1} to the second line, there is a boundary term that is discarded and that therefore cannot be recovered in this approach (this was stressed in \cite{Padmanabhan2004}).

The claim that general relativity is off-shell constructible would be precisely that the set $\{\Gamma_{(k)}\}_{k\geq3}^\infty$ is uniquely determined from the knowledge of $\Gamma_{(2)}$. Indeed, $\Gamma_{(2)}$ can be used to determine a Noether current associated with translation invariance, the so-called canonical stress-energy tensor. This Noether current can be used in order to couple the field $h_{ab}$ to itself, which introduces nonzero coefficients $\Gamma_{(3)}$ and might allow the unique determination of the latter. This procedure can be then applied recursively. We can identify a number of issues with this procedure:
\begin{itemize}
\item[a)]{\emph{Field redefinitions:} these redefinitions change the form of the action without modifying the actual physics of the theory. In other words, there are different sets of coefficients $\{\Gamma_{(k)}\}_{k\geq3}^\infty$ that nevertheless lead to the same on-shell behavior (it is also worth keeping in mind that the values of these coefficients depend on the choice of gauge-fixing). It is not clear how and why this procedure would be able to pick a specific off-shell realization (in particular, the one that corresponds precisely to the Einstein-Hilbert action).}
\item[b)]{\emph{Non-uniqueness:} a specific set of values for $\Gamma_{(2)}$, and a particular Noether current derived from them, leads to general relativity \cite{Deser1969}. However, the coefficients $\Gamma_{(3)}$ cannot be uniquely determined from $\Gamma_{(2)}$, as Noether currents are not uniquely defined (it is always possible to add identically conserved pieces to these currents). This feature shows up at every step of the iterative procedure, thus leading to a cumulative non-uniqueness \cite{Padmanabhan2004,Barcelo2014} (see also the explicit discussion of this issue in App. \ref{eq:uniq}). Additional arguments would be needed in order to discard these other solutions, but it is not clear whether these arguments exist.}
\item[c)]{\emph{Higher-derivative interactions:} there is no reason to consider from the beginning an ansatz such as the one on the second line of Eq. \eqref{eq:cons1}, containing interactions that are only quadratic on the derivatives of the field $h_{ab}$ (let us recall that this approach does not assume by construction diffeomorphism invariance from the beginning, so that this symmetry cannot be used in order to reduce the number of possible interaction terms in the initial ansatz). This feature has been always put by hand \cite{Padmanabhan2004,Deser2009} without further justification. This issue is entangled with the nature of the identically conserved pieces that can be added to the stress-energy tensor. Disregarding higher-derivative interactions becomes more questionable due to the existence of higher-derivative theories with just two degrees of freedom that reduce to gravitons at the linear level \cite{Lin2017,Carballo-Rubio2018} (see also \cite{Aoki2018,Lin2018}).}
\end{itemize}
Aside from these issues, there are two additional points that, while minor, serve nevertheless to illustrate the difference with respect to the on-shell approach discussed below:
\begin{itemize}
\item[d)]{\emph{Prior knowledge of general relativity:} attempts at deriving the Einstein-Hilbert action have been typically contaminated with the knowledge of the desired outcome. This was thoroughly discussed in \cite{Padmanabhan2004}, including using the Hilbert prescription to obtain the stress-energy tensor which, however, is not necessary as one could equally use the canonical stress-energy tensor derived using only flat-spacetime notions, as emphasized in \cite{Barcelo2014}. Nevertheless, it would be desirable to find approaches that make even more clear that there are no traces of geometric notions associated with curved spacetimes.}
\item[e)]{\emph{The root of constructibility:} off-shell approaches fail to justify what makes general relativity special so that it is off-shell constructible. The work \cite{Deser1969} strongly suggests that off-shell constructibility is associated with gauge invariance. However, the on-shell techniques described below present a different take on this issue, as on-shell constructibility is a much more general feature of interacting quantum field theories.}
\end{itemize}
Let us now turn our attention to the on-shell description in terms of scattering amplitudes. A given set $\{\Gamma_{(k)}\}_{k=2}^\infty$ can be used in order to determine the set of amplitudes $\{A_n(1^{h_1}2^{h_2}...\,n^{h_n})\}_{n=3}^\infty$. In fact, the subset of coefficients in the action with $k\leq n$ uniquely determines the $n$-point amplitudes:
\begin{equation}\label{eq:gammatoan}
\{\Gamma_{(k)}\}_{k=2}^{n}\rightarrow A_n(1^{h_1}2^{h_2}...\,n^{h_n}).
\end{equation}
More explicitly, $\{\Gamma_{(k)}\}_{k=2}^{n}$ permits us to write down the relevant Feynman rules, that can be then used in order to calculate the amplitudes $A_n(1^{h_1}2^{h_2}...\,n^{h_n})$. The inverse statement is not true, in particular due to point (a) above regarding field redefinitions.

Working with the set $\{A_n(1^{h_1}2^{h_2}...\,n^{h_n})\}_{n=3}^\infty$ instead permits us to avoid the issues associated with the off-shell approach, that is focused on $\{\Gamma_{(k)}\}_{k=2}^\infty$. It is straightforward to see that point (a) is avoided, as the S-matrix is invariant under field redefinitions \cite{Chisholm1961,Kamefuchi1961,tHooft1973}. Point (d) is trivially dealt with, as the discussion is now framed in terms of the language of standard quantum field theory in flat spacetime. Understanding how the situation with respect to points (b), (c) and (e) may change requires that we recall some results obtained using modern techniques for the calculation of amplitudes, including the notion of constructibility defined in the introduction.

General relativity has been shown to be constructible via the BCFW relations \cite{Britto:2004,Britto:2005,Benincasa2007,ArkaniHamed2008}. For the purposes of this paper, we just need to recall that $n_\star=3$ in this case, meaning that 3-point amplitudes are enough to determine all the remaining $n$-point amplitudes. This procedure is unique once the 3-point amplitudes are fixed. This strongly alleviates point (b) regarding the non-uniqueness of the off-shell approach that was present at every step, making this issue more manageable, as non-uniqueness is clearly confined to the seed $A_3(1^{h_1}2^{h_2}3^{h_3})$. Regarding point (c), the recursive derivation based on the BCFW relations is also limited by construction to interactions that are at most quadratic in the derivatives of the field $h_{ab}$ (see \cite{Carballo-Rubio2018} for an explicit discussion). At first sight this may suggest that the on-shell approach would suffer from the same drawback of the off-shell approach. However, in this paper we see that, using soft theorems, it is possible to deal with higher-derivative interactions. Regarding the last point (e), the on-shell approach shows that constructibility is a more general feature of (effective) quantum field theories \cite{Cohen2010,Cheung2015,Cheung2015b,Cheung2016}.

One may say that a shortcoming of the on-shell approach is that it does not allow us to obtain the Einstein-Hilbert action uniquely. That is, once the set of amplitudes $\{A_n(1^{h_1}2^{h_2}...\,n^{h_n})\}_{n=3}^\infty$ is obtained, we know that the Einstein-Hilbert action is one of the possible off-shell realizations leading to these amplitudes. However, as stressed above, it is not possible to carry out the inverse procedure and evaluate the action in a unique way. But it is important to recall that the off-shell approach suffers from the very same issue, as discussed above. Hence, we can conclude that the on-shell approach is more convenient in the sense that it offers a number of improvements with respect to the off-shell approach, without any actual drawbacks.

\section{Our main statement}\label{sec:statement}

In this section, we formulate a general set of criteria that must be met in order to determine completely the tree-level S-matrix of a given gravitational action with two local degrees of freedom, showing its equivalence with general relativity on asymptotically flat spacetimes at all orders in perturbation theory. This set of criteria is general enough to include higher-derivative interactions and is formulated without relying on particular off-shell symmetries (such as diffeomorphism invariance). Aside from the conditions below, we assume the standard requisites of locality and unitarity \cite{Conde2013,Benincasa2013}, and we work in $D=4$ dimensions.
 
Let us assume that there exists a gravitational action such that:
\begin{itemize}
\item[A)]{Describes two degrees of freedom that, at the linear level, correspond to massless gravitons;}
\item[B)]{Has the same 3-point amplitudes as general relativity;}
\end{itemize}
then, it follows that
\begin{itemize}
	\item[1)]{The soft graviton theorem with the standard leading, subleading and\linebreak sub-subleading contributions is verified.}
\end{itemize}
Furthermore, if we also assume that:
\begin{itemize}
\item[C)]{The propagator behaves for large (off-shell) momentum as $T_{\mu\nu\rho\sigma}(p)/(p^{2})^{m/2+1}$, where $T_{\mu\nu\rho\sigma}(p)$ represents an arbitrary tensorial structure (perhaps Lorentz violating) containing $0\leq m\leq 4$ times the product of the momentum $p$;}
\item[D)]{$k$-point interaction vertices, with $k\geq 4$, have at most $I(k)\leq I_\star(k)=2(k-1)$ powers of momenta, while for $3$-point interaction vertices we demand that $I(3)< 4-m/3$;}
\end{itemize}
then,
\begin{itemize} 
	\item[2)]{The entire (tree-level) S-matrix is recursively constructible from the information encoded in the soft graviton theorem, and is therefore the same as in general relativity.}
\end{itemize}
The proof of 1 and 2 is provided in Sec. \ref{sec:proof} below.

\section{Derivation}\label{sec:proof}

\subsection{Soft theorem from a double complex shift\label{sec:softth}}

We will follow \cite{Elvang2016} for the derivation of the soft graviton theorem (see also \cite{Laddha2017}). That the soft standard soft theorem applies with no modification to the theories satisfying (A-B) above is, in fact, a direct consequence of the discussion in \cite{Elvang2016}. Let us start with a brief review of the steps in this derivation, up to the point in which it is possible to formulate the result we want to highlight. Let us introduce the double complex deformation
\begin{align}\label{eq:3shift}
|\hat{s}\rangle&=\epsilon|s\rangle-z|X\rangle,\nonumber\\
|\hat{i}]&=|i]-\epsilon\frac{\langle js\rangle}{\langle ji\rangle}|s]+z\frac{\langle jX\rangle}{\langle ji\rangle}|s],\nonumber\\
|\hat{j}]&=|j]-\epsilon\frac{\langle is\rangle}{\langle ij\rangle}|s]+z\frac{\langle iX\rangle}{\langle ij\rangle}|s].
\end{align}
Under this deformation, a given amplitude $A_{n+1}$ becomes a function of two complex variables $z$ and $\epsilon$, $\hat{A}_{n+1}(z,\epsilon)$. The arguments below are formulated in the region of $\mathbb{C}^2$ defined by $z\ll\epsilon$. The limit $\epsilon\rightarrow0$ (with $z\rightarrow0$ as well, such that $z\ll\epsilon$ always) corresponds then to the (holomorphic) soft limit in which the momentum of the particle $s$ vanishes. The arguments below do not depend on the particular choice of particles $i$ and $j$ inside the set $k\in[1,n]$ and the arbitrary spinor $|X\rangle$ \cite{Elvang2016}.

The possible poles of $\hat{A}_{n+1}(z,\epsilon)$ can come only from internal momenta becoming on-shell, which means that this function is meromorphic, given that locality implies that these poles should be always associated with propagators of the form
\begin{equation}\label{eq:dshpoles}
\frac{1}{(\chi\hat{p}_s+\hat{P}_J)^2},
\end{equation}
where $\hat{P}_J=\sum_{l\in J}\hat{p}_l$, $J\subset[1,n]$ and $\chi\in\{0,1\}$. There are two kinds of poles. The first class of poles, $\{\epsilon_k\}$, are linear in $z$, and their location approaches the origin as $z\rightarrow0$. Poles in the second class, $\{\bar{\epsilon}_m\}$, satisfy $\lim_{z\rightarrow0}\bar{\epsilon}_m=\bar{\epsilon}_m^0=\pazocal{O}(\epsilon^0)$. We can see from Eq. \eqref{eq:dshpoles} that there are $k\in[1,n]$ poles in the first class that are, moreover, simple poles, for which $\chi=1$ and $J$ has just a single element:
\begin{equation}
(\hat{p}_s+\hat{p}_k)^2=\langle\hat{s}k\rangle[s\hat{k}]=(\epsilon-\epsilon_k)\langle sk\rangle[sk],\qquad \epsilon_k=z\frac{\langle Xk\rangle}{\langle sk\rangle}.
\end{equation}
It is not difficult to show that, if $\chi=0$ or $J$ has more than one element, the corresponding poles must be in the second class.

Given that $\hat{A}_{n+1}(z,\epsilon)$ is meromorphic, we can always write it as
\begin{equation}
\hat{A}_{n+1}(z,\epsilon)=\sum_{k=1}^n\frac{\mbox{Res}_{\epsilon=\epsilon_k}\hat{A}_{n+1}}{\epsilon-\epsilon_k}+D(z,\epsilon)+\pazocal{O}(\epsilon^0),
\end{equation}
where $D(z,\epsilon)$ is a meromorphic function containing all the poles in the second class. It is important to stress that the amplitude $\hat{A}_{n+1}(z,\epsilon)$ may have a residue at $\epsilon=\infty$, but this contribution would be contained in the $\pazocal{O}(\epsilon^0)$ part of the previous equation. Therefore, the soft limit $\epsilon\rightarrow0$ is insensitive to the existence of this residue at infinity or, equivalently, to the validity of the shift defined in Eq. \eqref{eq:3shift}.
 
Taking into account explicitly that $z\ll\epsilon$, we can write
\begin{equation}
\hat{A}_{n+1}(z\ll\epsilon,\epsilon)=\sum_{k=1}^n\frac{\mbox{Res}_{\epsilon=\epsilon_k}\hat{A}_{n+1}}{\epsilon}\left(1+\frac{\epsilon_k}{\epsilon}+...\right)+D(z\ll\epsilon,\epsilon)+\pazocal{O}(\epsilon^0).
\end{equation}
It is then clear that the divergent behaviour in the limit $\epsilon\rightarrow0$ is isolated in the first term on the right-hand side of the previous equation. Hence, we can ignore the second term, which will be $\pazocal{O}(\epsilon^0)$ in the soft limit, and write simply
\begin{equation}\label{eq:softder1}
\hat{A}_{n+1}(z\ll\epsilon,\epsilon\ll1)=\sum_{k=1}^n\frac{\mbox{Res}_{\epsilon=\epsilon_k}\hat{A}_{n+1}}{\epsilon}\left(1+\frac{\epsilon_k}{\epsilon}+...\right)+\pazocal{O}(\epsilon^0).
\end{equation}
This equation can be now used in order to obtain soft theorems, as discussed in \cite{Elvang2016}. In particular, it can be used in order to show that the certain theories of gravity satisfy the standard soft graviton theorem. 

In order to do so, we need the form of the residues $\mbox{Res}_{\epsilon=\epsilon_k}\hat{A}_{n+1}$, which is fixed by the condition (A) and unitarity. Unitarity implies the factorization of amplitudes around simple poles $\epsilon=\epsilon_k$ that come only from 2-particle channels (e.g., \cite{Eden1966,Conde2013}), so that the amplitude factorizes into a product of a 3-point amplitude and an $n$-point amplitude,
\begin{equation}\label{eq:softder2}
\mbox{Res}_{\epsilon=\epsilon_k}\hat{A}_{n+1}=\sum_{h_k}\frac{\hat{A_3}(z,\epsilon_k)\hat{A}_n(z,\epsilon_k)}{\langle sk\rangle[sk]},
\end{equation}
where $\hat{A_3}(z,\epsilon_k)$ is a 3-point on-shell amplitude and the sum is performed over the helicity $h_k$ of the internal particle that goes on-shell. Soft theorems can be calculated directly from the multiplicative factor in front of $\hat{A}_n(z,\epsilon_k)$ and the Laurent expansion around $z=0$ of the latter \cite{Elvang2016}.

Condition (B) implies that 3-point on-shell amplitudes are identical to those of general relativity. From Eqs. \eqref{eq:softder1} and \eqref{eq:softder2} above, it follows that the soft graviton theorem could be modified only through modifications of the 3-point on-shell amplitudes. Hence, condition (B) fixes the form of the soft graviton theorem to be the standard one.

For completeness, let us write explicitly the form of the (positive helicity) soft graviton theorem that can be directly obtained from Eqs. \eqref{eq:softder1} and \eqref{eq:softder2} by taking the following steps (we refer the reader to \cite{Elvang2016} for a more detailed discussion). The Laurent expansion of Eq. \eqref{eq:softder1} around $z = 0$ gives a series of poles in $z$, the order of which is dependent on the theory under consideration. For gravitational theories with the same 3-point amplitudes as general relativity, the coefficients of both $z^{-2}$ and $z^{-1}$ terms must vanish on-shell (which leads to Weinberg's formulation of the equivalence principle \cite{Weinberg1964}), since the final amplitude should not have any poles in $z$ due to the requirement of locality. The remaining finite $z^0$ piece of the expansion is exactly the (positive-helicity) soft graviton theorem,
\begin{align}\label{eq:softth}
A_{n+1}&=(...,\epsilon|s\rangle,|s])=\left(\frac{1}{\epsilon^3}\pazocal{S}^{(0)}+\frac{1}{\epsilon^2}\pazocal{S}^{(1)}+\frac{1}{\epsilon}\pazocal{S}^{(2)}\right)A_n+\pazocal{O}(\epsilon^0).
\end{align}
The $\pazocal{O}(\epsilon^0)$ terms are not universal, or their form is not known. On the other hand, the quantities $\pazocal{S}^{(k)}$ are operators that act on the amplitude $A_n$ in the previous equation, and are given by \cite{Elvang2016,Cachazo2014}
\begin{align}
\pazocal{S}^{(0)}&=\sum_{a=1}^n\frac{[sa]}{\langle sa\rangle}\frac{\langle xa\rangle\langle ya\rangle}{\langle xs\rangle \langle ys\rangle},\nonumber\\
\pazocal{S}^{(1)}&=\frac{1}{2}\sum_{a=1}^n\frac{[sa]}{\langle sa\rangle}\left(\frac{\langle xa\rangle}{\langle xs\rangle}+\frac{\langle ya\rangle}{\langle ys\rangle}\right)\mathscr{D}_{sa},\nonumber\\
\pazocal{S}^{(2)}&=\frac{1}{2}\sum_{a=1}^n\frac{[sa]}{\langle sa\rangle}\mathscr{D}_{sa}^2,
\end{align}
where $\mathscr{D}_{sa}=|s]_b\partial_{|a]_b}$. This derivative operator comes directly from the expansion of $\hat{A}_n(z)$ in $z$ in Eq. \eqref{eq:softder2}. The $\pazocal{S}^{(2)}$ term was first identified in \cite{Cachazo2014}, and it will be of essential importance for our discussion in Sec. \ref{sec:allline}.

Before ending this section, it is interesting to recall that the information encoded in Eq. \eqref{eq:softth} is enough to fix the $n=4$ and $n=5$ maximal helicity violating (MHV) graviton amplitudes \cite{Cachazo2014}. Regarding the other amplitudes with $n=4$ and $n=5$ external gravitons, it is straightforward to show that condition (D) above guarantees they can be fixed either by means of a BCFW shift such as the one used in order to derive the auxiliary recursion relations in \cite{Benincasa2007} affecting all particles in the amplitudes $A_4(+,+,+,-)$ and $A_5(+,+,+,+,-)$, or a CSW shift \cite{Cachazo:2004} affecting all particles in the amplitudes $A_4(+,+,+,+)$ and $A_5(+,+,+,+,+)$. These shifts fail however in order to deal with higher values of $n$, unless the numbers of powers of momenta are more restricted than our condition (D); see for instance \cite{Carballo-Rubio2018} for an explicit discussion. Hence, we can conclude that the $n=4$ and $n=5$ amplitudes in the theories satisfying (A-B) are the same as in general relativity, but also that we will need to consider a different strategy to deal with $n\geq6$ amplitudes. Let us also note that there is a similar relation as Eq. \eqref{eq:softth} that is valid for negative-helicity particles, and which can be obtained following the same steps but taking the soft limit as $\epsilon|s]$.

\subsection{All-line shift and on-shell recurrence relations from the soft theorem \label{sec:allline}}

Having established that the the standard soft graviton theorem is satisfied, let us now exploit this information in order to construct the scattering amplitudes for $n\geq6$. Let us consider the effect of the following all-line shift on $A_n$ for $n\geq 6$ (see \cite{Cheung2015b,Luo2015} for previous applications of this shift), 
\begin{equation}\label{eq:defshift}
\hat{p}_i=p_i(1-a_iz),\qquad i\in[1,n].
\end{equation}
The momentum conservation constraint, 
\begin{equation}\label{eq:zcons}
\sum_{i=1}^n\hat{p}_i=z\sum_{i=1}^na_i p_i=0,
\end{equation}
has a nontrivial solution (the trivial solution would correspond to all the coefficients $a_i$ being equal) for $z\neq0$ only for $n>D+1$, where $D$ is the dimension of spacetime. For $D=4$, there is a nontrivial solution to Eq. \eqref{eq:zcons} only for $n\geq6$, which are nevertheless the only remaining cases we have to deal with. Particular solutions to Eq. \eqref{eq:zcons} were given in \cite{Cheung2015b}, taking into account that only $D$ of the momenta can be linearly independent, as explained in the following. Let us consider a subset $K\subset\{i\}_{i=1}^n$ of $D+1$ indices chosen from the indices labelling the external particles, that inherits the order of $\{i\}_{i=1}^n$. Given $j\in K$, we can then define
\begin{equation}
a_j=\frac{(-1)^j}{D!}\varepsilon^{r_1r_2...r_D}\varepsilon^{b_1b_2...b_D}(p_{r_1})_{b_1} (p_{r_2})_{b_2}...\ (p_{r_D})_{b_D},
\end{equation}
where $r_s$ takes all the values in $K\backslash\{j\}$ for every $s\in[1,D]$, and $\varepsilon^{r_1r_2...r_D}$ is the $D$-dimensional Levi-Civita symbol. This provides a solution of Eq. \eqref{eq:zcons} such that the coefficients $\{a_j\}_{j\in K}$ are nonzero while the remaining coefficients, with indices on the set $\{i\}_{i=1}^n\backslash K$, are identically vanishing. Given that Eq. \eqref{eq:zcons} is linear, we can consider linear combinations of these solutions for different choices of $K\subset\{i\}_{i=1}^n$ in order to generate solutions with all coefficients $\{a_i\}_{i=1}^n$ nonzero (for instance, for $n=6$ we need to consider at least two of these subsets of indices that must be, of course, different). For generic configurations of the external momenta, these coefficients will moreover be distinct. In summary, in the following we will always limit our discussion to the case in which all coefficients $a_i$ are distinct and nonzero. One could allow for some of these coefficients to be the same, which would correspond to multi-soft limits, although we do not need to consider these cases for the present discussion.

The shift in Eq. \eqref{eq:defshift} is such that the momentum of the particle $s$ becomes soft for $z=z_s=1/a_s$. We can characterize the way in which the soft limit is approached defining $z=(1-\epsilon)/a_s$, where $\epsilon\ll1$ and the definition is valid in the vicinity of each $z_s$. We can then deform the spinor-helicity variables of the soft momentum holomorphically, such that the soft limit $\epsilon\rightarrow0$ is taken either as in Eq. \eqref{eq:softth}, namely as
\begin{equation}\label{eq:holo}
|\hat{s}\rangle=\epsilon|s\rangle,
\end{equation}
or antiholomorphically as
\begin{equation}\label{eq:antiholo}
|\hat{s}]=\epsilon|s].
\end{equation}
The helicity of each external particle will determine which of these is chosen in each case \cite{Cheung2015}, in order to obtain the best possible bound on the behavior with large $z$ in Sec. \ref{sec:val2}.

The shift in Eq. \eqref{eq:defshift} defines a complexification $\hat{A}_n(z)$ of $n$-point amplitudes, such that the physical amplitudes are given by $A_n=\hat{A}_n(0)$. Let us define the function of complex variable\footnote{Let us remark that our definition of $f_n(z)$ does not contain additional multiplicative factors of the form $1/(1-a_iz)^\sigma$ for $\sigma>0$, that are typically considered when exploiting soft theorems in order to derive on-shell recursion relations \cite{Cheung2015b,Luo2015}. The reason is that, in theories satisfying the soft theorem \eqref{eq:softth}, this would originate additional poles coming from the $\pazocal{O}(\epsilon^0)$ pieces, thus preventing the recursive evaluation of scattering amplitudes.}
\begin{equation}
f_{n}(z)=\frac{\hat{A}_n(z)}{z}.
\end{equation}
This function exhibits three different kinds of singularities. Aside from the trivial $z=0$ pole, the remaining singularities are inherited from the singularity structure of the physical amplitude $A_n$, resulting in factorization poles corresponding to internal momenta going on-shell and around which the amplitude factorizes into the product of two sub-amplitudes, and soft poles at $z=1/a_i$ due to $\hat{p}_i$ becoming soft. Factorization poles arise whenever the momentum of an external particle and the vector arising from the linear combination of the momenta of other external particles become collinear; the particular cases when the momenta of two external particles become collinear are known as collinear limits, while the remaining cases that involve at least three particles are known as multi-particle channels \cite{Benincasa2013,Conde2013}. Soft poles can be understood as a particular case of a collinear limit in which the two collinear momenta vanish. However, one must keep in mind that the additional condition of these momenta vanishing implies that the factorization property into two subamplitudes is lost, and also modifies the degree of divergence around these singularities \cite{Conde2013}.

It is straightforward to see that, for the shift \eqref{eq:defshift}, all factorization poles in the complex variable come from multi-particle factorization poles in $z$ \cite{Luo2015}, so that collinear limits yield no factorization poles. Indeed, the singular structure of the corresponding propagator in the collinear case is proportional to
\begin{equation}
\frac{1}{\hat{p}_i\cdot\hat{p}_j}=\frac{1}{p_i\cdot p_j(1-a_iz)(1-a_jz)}.    
\end{equation}
Hence, for this particular shift, collinear limits yield only soft poles in $z$. This will be of importance later.

The possible existence of soft poles in theories with massless particles implies that the shift \eqref{eq:defshift} would be generally useless (from the perspective of constructing the tree-level S-matrix) unless the soft behavior, namely whether or not there are soft poles and their corresponding degree and residues, is completely determined.

The existence of poles at $z=1/a_i$ is guaranteed from the soft theorem in Eq. \eqref{eq:softth}. Other factorization channels lead to poles arising from a quadratic equation, the two roots of which will be denoted as $\{z_I^\pm\}$ \cite{Cheung2015b,Luo2015}. There are no additional poles, as all poles of $\hat{A}_n(z)$ must be associated either with one of the internal momenta becoming on-shell or one of the external momenta becoming soft following our assumptions \cite{Conde2013}.

Let us assume that, for $|z|\rightarrow\infty$, the complex amplitude $\hat{A}_n(z)$ satisfies
\begin{equation}
\hat{A}_n(z)\propto z^{\delta(n)},
\end{equation}
for some integer (that may depend on $n$)
\begin{equation}\label{eq:validity}
\delta(n)<0.
\end{equation}
Integrating $f_{n}(z)$ on a contour $\gamma$ that encloses all its poles, taking the contour to $z\rightarrow\infty$, and applying Cauchy's residue theorem, permits us to write
\begin{equation}\label{eq:residueth}
A_n=-\sum_{I}\mbox{Res}_{z=z_I^\pm}\frac{\hat{A}_n(z)}{z}-\sum_{i=1}^n\mbox{Res}_{z=1/a_i}\frac{\hat{A}_n(z)}{z}.
\end{equation}
The left-hand side of the previous equation contains the physical $n$-point amplitude of real momenta $A_n$. The right-hand side contains the contributions from the $z\neq0$ poles. Let us stress that there is no contribution from the residue at infinity as we are assuming that Eq. \eqref{eq:validity} holds. We will determine in Sec. \ref{sec:val2} the situations in which this is indeed verified, which will lead to condition (D).

We have separated the two kinds of contributions from the poles of $\hat{A}(z)$ in Eq. \eqref{eq:residueth}. The poles in the first term of the right-hand side correspond to internal momenta going on-shell. Due to the factorization properties of scattering amplitudes \cite{Elvang2015}, it follows that this term can be written as a sum of products of lower-point amplitudes $A_{k<n}$ evaluated on complex momenta, which means that this term is constructible in a recursive manner \cite{Cheung2015b,Luo2015}:
\begin{equation}\label{eq:1stterm}
-\sum_{I}\mbox{Res}_{z=z_I^\pm}\frac{\hat{A}_n(z)}{z}=\sum_I\frac{1}{P_I^2}\frac{A_{\rm L}(z_I^-)A_{\rm R}(z_I^-)}{1-z_I^-/z_I^+}+(z_I^-\leftrightarrow z_I^+),
\end{equation}
where the sum must be taken over all possible factorization channels and values of the helicity of the internal particle with momentum $\hat{P}_I$ becoming on-shell.

The second contribution on the right-hand side of Eq. \eqref{eq:residueth} always arises due to the particular form of the shift \eqref{eq:defshift}, and explores the soft singular behavior of on-shell amplitudes. We can now use the results of Sec. \ref{sec:softth}, namely that the graviton soft theorem fully characterizes the singular soft behavior, so that we can write
\begin{align}\label{eq:softcont}
&-\sum_{i=1}^n\mbox{Res}_{z=1/a_i}\frac{\hat{A}_n(z)}{z}\nonumber\\
&=-\sum_{i=1}^{n_+}\mbox{Res}_{z=1/a_i}\frac{1}{z}\left\{\frac{1}{(1-a_iz)^3}\pazocal{S}^{(0)}+\frac{1}{(1-a_iz)^2}\pazocal{S}^{(1)}+\frac{1}{(1-a_iz)}\pazocal{S}^{(2)}\right\}\hat{A}_{n-1}(z)+...
\end{align}
We have only written the contribution coming from positive-helicity particles (we have also chosen without loss of generality a particular ordering of the particles), and the ellipsis indicates the presence of a structurally equivalent term but for negative-helicity particles. Hence, also this contribution is recursively constructible which, together with Eq. \eqref{eq:1stterm}, permits us to state that the complete $n$-point amplitudes $A_{n}$ are constructible from $A_{n-1}$. It is worth emphasizing that constructing these amplitudes using the all-line shift \eqref{eq:defshift} is only possible because all leading, subleading and sub-subleading terms in the soft graviton theorem are under control. In other words, without the identification of the term proportional to $\pazocal{S}^{(2)}\hat{A}_{n-1}(z)$ in \cite{Cachazo2014} (and the fact that it is completely determined by the amplitude $A_{n-1}$), it would have been impossible to construct the scattering amplitudes of generic gravitational theories using the shift \eqref{eq:defshift}. In particular, this seems to be the reason why this shift was not used in order to determine that general relativity is constructible at the time in which the original proofs of this statement were provided \cite{Benincasa2007,ArkaniHamed2008}. 

\subsection{Validity of the all-line shift \label{sec:val2}}

We have shown in Sec. \ref{sec:softth} that the standard soft graviton theorem is satisfied while, in Sec. \ref{sec:allline}, we have discussed that an all-line shift would allow us to construct scattering amplitudes in a recursive manner from the information encoded in the soft theorem. This constructibility condition relies on the validity of the shift, namely Eq. \eqref{eq:validity}, being satisfied. In this section we show that this condition is satisfied under the conditions (C-D) in Sec. \ref{sec:statement}. In order to do so, we need to obtain suitable bounds on the behavior of $\hat{A}_n(z)$ for $|z|\rightarrow\infty$. The simplest bound on the behavior of the complex amplitudes with $z$ one can obtain follows from the analysis of individual Feynman diagrams (see, e.g., \cite{Elvang2015} for an extended discussion).

Given a particular helicity arrangement, we make the optimal choice regarding polarization vectors. This means that for gravitons with positive helicity we will use holomorphic shift \eqref{eq:holo}, while for gravitons with negative helicity we use the antiholomorphic shift \eqref{eq:antiholo}. Hence, the leading contribution from the polarization vectors of an $n$-point amplitude is $(z^{-2})^n$. 

Each propagator contributes with a $z^{-2}$ factor, which follows directly from condition (C). On the other hand, let us assume that $k$-point interaction vertices display a leading behavior $z^{I(k)}$. Let us start considering individual Feynman diagrams with vertices of the same valence, which display the asymptotic behavior
\begin{equation}\label{eq:bound}
\hat{A}_n(z)\propto (z^{-2})^n(z^I)^{(n-2)/(k-2)}(z^{-2})^{(n-k)(k-2)}.
\end{equation}
The validity of the shift \eqref{eq:defshift} implies then that
\begin{equation}
\delta(n)\leq -2n+I\frac{n-2}{k-2}-2\frac{n-k}{k-2}<0.
\end{equation}
This equation must be satisfied for all values of $n\geq 6$ such that $n\geq k\geq3$, which leads to condition (D):
\begin{equation}\label{eq:validitycond}
I(k)\leq I_\star (k)=2(k-1).
\end{equation}
This equation represents at this stage of the discussion a necessary condition, given that we have only considered a particular class of diagrams in order to derive it. However, under this necessary condition, changing the valence of the vertices inside specific subdiagrams with $m\leq n$ legs does not modify the restriction of the bound \eqref{eq:bound} to these subdiagrams, that is always proportional to $z^{-2}$. Hence, Eq. \eqref{eq:validitycond} is in fact necessary and sufficient. One can check that this is consistent with the discussion of the same all-line shift in \cite{Elvang2018}. Besides of the all-line shift being valid, theories in which $I(k)$ does not saturate the inequality in Eq. \eqref{eq:validitycond} can display bonus relations in the sense of \cite{ArkaniHamed2008b,Spradlin2008}.

On the other hand, there may be vertices that vanish when one of their legs is on-shell (an example of this kind of theory is provided in Sec. \ref{sec:minmod}). These vertices can display a leading behavior $z^{J(k)}$, where $J$ is determined taking into account the number of external (with at least one external leg attached to them) and internal vertices in a given diagram. Let us recall that the number of internal vertices is constrained by
\begin{equation}
i(k)\leq \frac{n}{(k-1)(k-2)}-\frac{2}{k-2}.
\end{equation}
Assuming that $J>I$, the worst-case scenario is the one in which the inequality above is saturated. One obtains then
\begin{equation}
J(k)\leq J_{\star}(k)=2(k-1)^2-(k-2)I(k).
\end{equation}
For $I=I_\star$, we see that $J_\star=I_\star$. Hence, vertices that vanish when at least one leg is on-shell cannot go beyond $J_\star=2(k-1)^2-I(k-2)$.

This finishes the proof, showing that all the theories with actions satisfying (A-D) in Sec. \ref{sec:statement} are constructible from the information encoded in the soft graviton theorem. Moreover, given that the soft theorem in these theories must take the same form as in general relativity, we conclude that the complete S-matrix of these theories must be the same as the S-matrix of general relativity.

\section{Applications}\label{sec:app}
%

\subsection{Constructibility of general relativity from soft theorems}\label{sec:grconst}

A straightforward application of the general result above is the case of general relativity itself. Of course, general relativity is known to be constructible using the Britto-Cachazo-Feng-Witten (BCFW) shift \cite{Benincasa2007,ArkaniHamed2008}. However, it is still interesting to understand whether the amplitudes of general relativity could be evaluated from the information encoded in soft theorems (in fact, a different proof of this statement has been recently presented in \cite{Rodina2018}). General relativity certainly satisfies the criteria (A-D) above, as its interactions are quadratic in the momenta ($I=2$), and is therefore constructible following the procedure described in this paper.

\subsection{Scattering amplitudes of unimodular gravity}\label{sec:ugconst}

Another interesting modification of general relativity is unimodular gravity, which is defined in terms of a different action that is invariant under transverse diffeomorphisms (and, in some formulations, Weyl transformations \cite{Alvarez2006,Carballo-Rubio2015}). Even though the classical field equations in vacuum are Einstein manifolds as in general relativity, a proof of its constructibility was lacking. Previous proofs of the constructibility of general relativity do not immediately apply to unimodular gravity due to the different off-shell behavior of the propagator which, as discussed in \cite{Alvarez2016}, must contain triple poles in $p^2$ in order to reproduce the standard Newtonian potential (this seems to have been missed in \cite{Burger2015}). In other words, the off-shell propagator contains a piece proportional to
\begin{equation}
\frac{p_{\mu}p_{\nu}p_{\rho}p_{\sigma}}{p^6}.
\end{equation}
Under a BCFW shift this propagator diverges as $z$, hence the original proof of constructibility for general relativity (that rests on the more standard $1/z$ scaling of propagators) cannot be applied to unimodular gravity. Nevertheless, the knowledge of the classical field equations of the theory strongly suggests that the scattering amplitudes of unimodular gravity must enjoy the same properties as the ones in general relativity, and therefore that the failure of showing the constructibility of unimodular gravity stems only from technical limitations associated with this particular shift.

We can apply our general result discussed above in order to close the issue of the constructibility of the scattering amplitudes of unimodular gravity. Unitarity, enforced as the condition $S^\dagger S= \mathds{1}$ on the S-matrix, implies that scattering amplitudes factorize around simple poles when internal momenta become on-shell \cite{Eden1966,Conde2013}. Hence, the contributions from the off-shell pieces in the propagator containing higher-order poles in $p^2$ must necessarily cancel on-shell in order to guarantee unitarity. If this cancellation does not take place, it would follow that the tree-level S-matrix of unimodular gravity is not be unitary. This cancellation can be seen explicitly for some of the $n=4$ and $n=5$ amplitudes, that have been evaluated using spinor-helicity variables and shown to be equivalent to the corresponding amplitudes in general relativity \cite{Alvarez2016}.

The analysis in our paper permits to conclude that this equivalence extends to the complete S-matrix. Unimodular gravity satisfies the condition (A) as it is equivalent to the Fierz-Pauli theory at the linear level \cite{Alvarez2006}. (B) and (D) are also satisfied, as unimodular gravity shares the 3-point amplitudes with general relativity and its interaction vertices are quadratic in the momenta \cite{Alvarez2016,Burger2015}. On the other hand, its propagator behaves as $1/p^2$ for large (off-shell) momenta, thus satisfying (C), with $m=4$. We can then conclude that the tree-level S-matrix of unimodular gravity is either the same as in general relativity, or it is non-unitary in the sense that $S^+S=\mathds{1}$ does not hold, and that this follows necessarily from the basic principles and requirements in our general discussion.

\subsection{Minimally modified theories of gravity}\label{sec:minmod}

Let us consider the theories of modified gravity introduced in \cite{Lin2017} (see also \cite{Carballo-Rubio2018,Aoki2018,Lin2018}). All the known examples of these theories that are radiatively stable have a Lagrangian density of the form
\begin{equation}\label{eq:minlag}
\mathscr{L} = \sqrt{-g}\,G(\pazocal{K},\pazocal{R}),
\end{equation}
where $\pazocal{R}$ is the 3-dimensional Ricci scalar and $\pazocal{K}=K_{ij}K^{ij}-K^2$ is quadratic in the extrinsic curvature $K_{ij}$. On the other hand, $G(\pazocal{K},\pazocal{R})$ is a function that satisfies the constraints presented in \cite{Lin2017} such that the theory propagates two local degrees of freedom. However, on general grounds, we can consider a perturbative expansion of this Lagrangian around Minkowski spacetime, which leads to
\begin{equation}\label{key}
\frac{\pazocal{L}}{\sqrt{-g}} = c_{\pazocal{R}}\pazocal{R} + c_{\pazocal{K}}\pazocal{K} + c_{\pazocal{R}\pazocal{K}}\pazocal{RK} +  c_{\pazocal{R}^2}\pazocal{R}^2 + \cdots,\qquad c_{\pazocal{X}^m\pazocal{Y}^n} = \left.\frac{1}{n!}\frac{\partial^{m+n}\pazocal{L}}{\partial\pazocal{X}^n\partial\pazocal{Y}^n}\right|_{\pazocal{R} = \pazocal{K} = 0}.
\end{equation}
In this perturbative expansion, the quantities $\pazocal{K}$ and $\pazocal{R}$ are expanded as well as
\begin{equation}
\pazocal{K}=\pazocal{K}^{(2)}+\pazocal{K}^{(3)}+...\sim p^2h^2+p^2h^3+...,\qquad \pazocal{R}=\pazocal{R}^{(1)}+\pazocal{R}^{(2)}+...\sim p^2h+p^2h^2+...,
\end{equation}
where the superindex indicates the number of powers of $h_{ab}$ (the perturbation with respect to the flat metric $\eta_{ab}$), and the right-hand side in each of these equations indicates schematically the behavior of the different terms in momentum space. This family of theories is determined by the couplings $c_{\pazocal{X}^m\pazocal{Y}^n}$ (which must satisfy some constraints) in their perturbative expansion, with general relativity being included as a particular choice. In fact, we can see that the first two terms in the expansion are proportional (up to rescalings and a boundary term) to the 4-dimensional Ricci scalar. This means that the first part of the Lagrangian is nothing but the Fierz-Pauli Lagrangian, so that this theory satisfies condition (A).

Evaluating the $\pazocal{R}^2$ term gives enough information to derive the propagator \cite{Carballo-Rubio2018},
\begin{equation}\label{eq:minprop}
\pazocal{D}_{\mu\nu\rho\sigma}(p) = \pazocal{F}_{\mu\nu\rho\sigma}(p) - \frac{c}{1+cp^2/2}\delta^0_\mu\delta^0_\nu\delta^0_\rho\delta^0_\sigma,
\end{equation}
where $\pazocal{F}_{\mu\nu\rho\sigma}(p)$ is the usual propagator derived from the Fierz-Pauli action in the de Donder gauge. Hence, the propagator satisfies condition (C). On the other hand, the on-shell 3-point amplitudes were derived in \cite{Carballo-Rubio2018} and were found to be the same as in general relativity, so that (B) is satisfied.

The last condition (D) has to do with the behavior of the interaction vertices with the momenta. There are two kinds of interaction vertices in the theory with Lagrangian \eqref{eq:minlag}. The first class encompasses $k$-point vertices that are nonzero off-shell. These are of the form $\pazocal{K}^{(2)}[\pazocal{R}^{(1)}]^{k-2}$ or $\pazocal{R}^{(2)}[\pazocal{R}^{(1)}]^{k-2}$ and therefore have precisely $I=I_\star=2(k-1)$, thus satisfying (D). On the other hand, there are $k$-point vertices that vanish identically when one of their legs is on-shell. These have $J=2k\geq J_\star$. We can conclude that (D) is not satisfied. Hence, even if the use of soft theorems improves the situation, it is still not possible to prove that the S-matrix in these theories is the same as in general relativity with the arguments in this paper.

\subsection{Most general 3-point graviton amplitudes}\label{sec:gen}

Condition (B) in Sec. \ref{sec:statement} assumes that the 3-point amplitudes of gravitons take the form that is obtained in general relativity. However, it is straightforward to check that the proof of constructibility still holds if we relax condition (B) but keep (A), (C) and (D) unchanged. Hence, we devote this section to discuss in more detail the possible freedom that may be allowed when relaxing this condition.

It is well-known that on-shell 3-point amplitudes (of complex momenta) can be fixed completely from kinematical considerations and little group scaling (hence, ultimately, Lorentz invariance). There are two independent on-shell 3-point amplitudes, that we can choose to be $A_3(1^{+2}2^{+2}3^{+2})$ and $A_3(1^{+2}2^{+2}3^{-2})$ without loss of generality. If (A) holds, little-group scaling fixes these amplitudes to be
\begin{equation}\label{eq:ind3pt}
A_3(1^{+2}2^{+2}3^{-2})=\kappa\frac{[12]^6}{[13]^2[23]^2},\qquad A_3(1^{+2}2^{+2}3^{+2})=\zeta\kappa^5[12]^2[13]^2[23]^2,
\end{equation}
where $\zeta\in\mathbb{R}$ is a dimensionless constant and $\kappa^2=8\pi G$. If $\zeta=0$ we recover the 3-point amplitudes of general relativity.

One may be tempted to argue that all these deformations of general relativity, that form a one-parameter family, are constructible. However, on dimensional grounds we can see that a nonzero $A_3(1^{+2}2^{+2}3^{+2})$ must be associated with higher-derivative interactions (see also \cite{Bai2016}). Hence, any theory with $\zeta\neq0$ would fail to satisfy (D) and, therefore, is not constructible using the arguments provided in this paper. If a more powerful treatment allows to relax condition (D) (perhaps, removing it completely), then one would be able to find the constructible theory that would result from the 3-points above with $\zeta\neq0$. It is worth mentioning that there is a natural candidate to be associated with the outcome of this hypothetical procedure, namely the theory known as Einsteinian cubic gravity \cite{Bueno2016,Hennigar2016,Bueno2016b}, which contains a cubic term in the curvature with the right dimension to generate the nonzero amplitude $A_3(1^{+2}2^{+2}3^{+2})$ in Eq. \eqref{eq:ind3pt}.

The expectation of the existence of a more powerful treatment can be further motivated by considering field redefinitions in general relativity. As we have discussed in Sec. \ref{sec:grconst}, it is possible to construct the tree-level S-matrix of general relativity exploiting the graviton soft theorem. However, we also know that the S-matrix is invariant under (nonlinear) local field redefinitions, and so we can use this freedom to do a general redefinition of the form $h\longrightarrow h + \kappa^{p + q - 1}\bar{\nabla}^p h^q$. Under such a field redefinition, we find that the cubic interaction in the Einstein-Hilbert action \eqref{eq:cons1} changes as 
\begin{equation}
     h\bar{\nabla}^2 h \longrightarrow h\bar{\nabla}^2h + \kappa^{2(p+q-1)}h^q\bar{\nabla}^{p+2}h^q +...,
\end{equation}
where the ellipsis indicates terms that are better behaved for large momenta and hence ignored. Provided that $q>1$, this redefinition does not affect the propagator, but does introduce $(2q)$-point vertices containing $p+2$ derivatives. For a generic value of $p$, these additional vertices spoil the constructibility, since condition (D) in our criteria is, in general, no longer met after this field redefinition is performed. That general relativity is indeed constructible illustrates that the direct counting of derivatives in interaction vertices cannot encapsulate all the physics.

We think that it is interesting to keep studying whether higher-derivative gravitational theories with $I\geq 2k$ can be shown to be constructible using other arguments. A hypothetical proof of constructibility of scattering amplitudes from $n=3$ that succeeds at relaxing (D) would imply the existence of only two independent gravitational theories satisfying (A) and (C): general relativity and Einsteinian cubic gravity. Aside from this extension, it would also be interesting to study the interplay between our results and recent related works such as \cite{BeltranJimenez2018,Casadio2018}. Let us also mention for completeness that, for higher dimensional Yang-Mills operators (that appear as the first term in an $\alpha'$ expansion in bosonic string theory and in other low energy effective string actions \cite{TSEYTLIN1986391}) such as $F^3 = f^{abc}F^{a\nu}_\mu F^{b\rho}_\nu F^{c\mu}_\rho$, the soft theorems remain unchanged \cite{Bianchi:2015,Elvang2016}. This might lead us to expect that gravitational amplitudes formed via the KLT relations also will not spoil the soft theorems, which is not true: the possible KLT products of two higher-order Yang-Mills amplitudes contains contributions from both $R^3$ terms at order $\alpha'^2$ and $\phi R^2$ terms at order $\alpha'$. On the other hand, Einsteinian cubic gravity does not contain terms that spoil the soft theorems, but does modify the 3-point amplitudes as we have seen in Eq. \eqref{eq:ind3pt}. Higher-dimensional operators than the ones considered in Einsteinian cubic gravity cannot alter the 3-point amplitudes and, provided they only contain helicity-2 modes, also cannot spoil the soft theorems.

\section{Conclusions}

In this paper we have formulated a set of criteria to determine the on-shell equivalence of gravitational theories, also discussing the relation with previous work regarding the derivation of general relativity from the principles of special relativity. In particular, we have shown that using soft theorems leads to an improved treatment of higher-derivative interactions. While our main result can be applied to both general relativity and unimodular gravity, showing the equivalence of their S-matrices, a crucial point of failure for higher-derivative theories is condition (D). It may be possible to relax condition (D), in the same way that the information encoded in soft theorems has allowed us to go up to $I=I_\star=2(k-1)$ powers of momenta (for $k$-point interaction vertices) instead of simply $I=2$. However, this would require further improvement in the understanding of the large $z$ behaviour of the amplitudes. Our discussion illustrates that this is not merely a technical point but that it has important practical implications for the understanding of the possible equivalence of gravitational theories. This provides, in our opinion, a strong motivation for further research in this direction.

\appendix

\section{On the non-uniqueness in Deser's derivation}\label{eq:uniq}

For completeness, let us analyze in more detail the source of non-uniqueness in the derivation of the Einstein-Hilbert action following Deser's procedure \cite{Deser1969}, in order to justify the importance of point b) in our discussion in Sec. \ref{sec:grcons}. After Padmanabhan stressed this issue in \cite{Padmanabhan2004}, Deser argued in \cite{Deser2009} that it should be possible to deal with this non-uniqueness performing suitable field redefinitions. However, as pointed out in \cite{Barcelo2014}, it is only possible to do this at the lowest order in the iterative off-shell procedure, which means that field redefinitions are not enough in order to remove the ambiguities in this procedure. Here, we want to provide a thoroughly explicit illustration of this point.

Let us use the same schematic notation as in \cite{Padmanabhan2004,Deser2009}, in which the Fierz-Pauli equations \cite{Fierz1939} are written as
\begin{equation}
\pazocal{D}^{abcd}h_{cd}=0,
\end{equation}
where $\pazocal{D}^{abcd}$ is a certain second-order differential operator satisfying
\begin{equation}
\bar{\nabla}_a\pazocal{D}^{abcd}=0.
\end{equation}
The first iteration yields equation of the form
equations \cite{Fierz1939} are written as
\begin{equation}\label{eq:1storder}
\pazocal{D}^{abcd}h_{cd}=\lambda T_{(1)}^{ab}(h)+\lambda\Delta_{(1)}^{ab}(h),
\end{equation}
where $\lambda$ is a coupling constant, $T^{ab}_{(1)}(h)$ the stress-energy tensor that yields the correct coupling in general relativity, and $\Delta_{(1)}^{ab}(h)$ a superpotential (i.e., an identically conserved tensor). The claim in \cite{Deser2009} is that $\Delta_{(1)}^{ab}(h)$ can be removed by a field redefinition
\begin{equation}\label{eq:redef}
h_{ab}=\tilde{h}_{ab}+\lambda\Theta(\tilde{h}),
\end{equation}
where $\Theta_{ab}$ satisfies the equation
\begin{equation}
\pazocal{D}^{abcd}\Theta_{ab}(\tilde{h})=\Delta_{(1)}^{ab}(\tilde{h}+\lambda\Theta).
\end{equation}
This equation is well-posed, as both sides of it are identically conserved. However, this field redefinition implies that Eq. \eqref{eq:1storder} reads
\begin{equation}\label{eq:1storderb}
\pazocal{D}^{abcd}\tilde{h}_{cd}=\lambda T_{(1)}^{ab}(\tilde{h}+\lambda\Theta)=\lambda T_{(1)}^{ab}(\tilde{h})+\lambda^2W_{(1)}^{ab}(\tilde{h}).
\end{equation}
The second identity can be alternatively seen as the definition of the tensor $W_{(1)}^{ab}$. The contribution proportional to the latter is $\pazocal{O}(\lambda^2)$, and comes from the intrinsically nonlinear nature of the iterative procedure. This piece should be taken into account in the second iteration, in which the field equations read
\begin{equation}\label{eq:2ndorder}
\pazocal{D}^{abcd}\tilde{h}_{cd}=\lambda T_{(1)}^{ab}(\tilde{h})+\lambda^2 T_{(2)}^{ab}(\tilde{h})+\lambda^2W_{(1)}^{ab}(\tilde{h})+\lambda^2\Delta_{(2)}^{ab}(\tilde{h}).
\end{equation}
$\Delta^{ab}_{(2)}(\tilde{h})$ is again a superpotential and, therefore, can be dealt with at this order by another shift similar to the one in Eq. \eqref{eq:redef} but adding $\pazocal{O}(\lambda^2)$ terms. Let us focus our attention on $W_{(1)}^{ab}(\tilde{h})$, which is not identically conserved. In fact, it is straighftorward to show that
\begin{equation}
\lambda\bar{\nabla}_aW_{(1)}^{ab}(\tilde{h})=\bar{\nabla}_a\left[T_{(1)}^{ab}(h)- T_{(1)}^{ab}(\tilde{h})\right]=\pazocal{O}(\lambda),
\end{equation}
where the $\pazocal{O}(\lambda)$ terms in the last identity arise from the $\lambda\Delta_{(1)}^{ab}$ in Eq. \eqref{eq:1storder}. This implies that $W_{(1)}^{ab}$ is not conserved and, moreover, that this lack of conservation appears at $\pazocal{O}(\lambda^2)$ in Eq. \eqref{eq:2ndorder} and therefore cannot be pushed forward to the next interation that includes $\pazocal{O}(\lambda^3)$ terms. Hence, it is impossible to remove completely the term $\Delta_{(1)}^{ab}(h)$ doing a field redefinition, as the fact that the stress-energy tensor itself depends on $h_{ab}$ leads to an additional residue $W_{(1)}^{ab}$ that is not conserved. Similarly, trying to absorb the $\Delta_{(n)}^{ab}$ terms results into non-removable terms $W_{(n)}^{ab}$, which illustrates how this non-uniqueness manifests at every step in the off-shell procedure. 

\acknowledgments

The authors would like to thank Freddy Cachazo, Alfredo Guevara, Stefano Liberati and Dimitar Ivanov for useful discussions. RCR and NM are grateful for the hospitality of Perimeter Institute, where this work was conceived. This research was supported in part by Perimeter Institute for Theoretical Physics. Research at Perimeter Institute is supported by the Government of Canada through the Department of Innovation, Science and Economic Development and by the Province of Ontario through the Ministry of Research and Innovation. NM is supported by the South African Research Chairs Initiative of the Department of Science and Technology and the National Research Foundation of South Africa. Any opinion, finding and conclusion or recommendation expressed in this material is that of the authors and the NRF does not accept any liability in this regard.

\bibliography{refs}	

\end{document}